\documentclass[]{interact}

\usepackage{epstopdf}% To incorporate .eps illustrations using PDFLaTeX, etc.
\usepackage[caption=false]{subfig}% Support for small, `sub' figures and tables

\usepackage[numbers,sort&compress]{natbib}% Citation support using natbib.sty
\bibpunct[, ]{[}{]}{,}{n}{,}{,}% Citation support using natbib.sty
\makeatletter% @ becomes a letter
\def\NAT@def@citea{\def\@citea{\NAT@separator}}% Suppress spaces between citations using natbib.sty
\makeatother% @ becomes a symbol again

\theoremstyle{plain}% Theorem-like structures provided by amsthm.sty
\newtheorem{theorem}{Theorem}[section]
\newtheorem{lemma}[theorem]{Lemma}

\theoremstyle{definition}

\theoremstyle{remark}

\usepackage{bm, comment}
\usepackage{amsfonts}
\usepackage{mathrsfs} %use \mathscr for \cal
\usepackage{hyperref}
\usepackage[final]{changes}  %
\newcommand{\tr}{{\mbox{\tiny{$\mathrm{T}$}}}}

\newcommand{\B}{\mathrm{B}}
\newcommand{\E}{\mathrm{E}}
\newcommand{\K}{\mathrm{K}}
\newcommand{\hist}{\mathrm{H}}
\newcommand{\Sd}{\mathrm{SD}}
\newcommand{\Var}{\mathrm{var}}

\newcommand{\sfrac}[2]{\mbox{$\frac{#1}{#2}$}}
\newtheorem{assump}{}

\newtheorem{rem}{\bf Remark}

\begin{document}

\title{Bernstein Polynomial Model for Nonparametric Multivariate Density}
\author{
\name{Tao Wang\textsuperscript{a} and Zhong Guan\textsuperscript{b}\thanks{CONTACT Zhong Guan. Email: zguan@iusb.edu}}
\affil{\textsuperscript{a}School of Mathematical Sciences, Harbin Normal University, Harbin, China;
\textsuperscript{b}Department of Mathematical Sciences, Indiana University South Bend, South Bend, Indiana, USA}
}

\maketitle
\begin{abstract}
In this paper, we study the Bernstein polynomial model for estimating the multivariate distribution  functions and densities with bounded support. As a mixture model of
multivariate beta distributions, the  maximum (approximate) likelihood estimate can be obtained using EM algorithm. A change-point method of choosing optimal degrees of the proposed Bernstein polynomial model is presented. Under some conditions the optimal rate of convergence in the mean $\chi^2$-divergence of new density estimator is shown to be nearly parametric.
The method is illustrated by an application to a real data set.
Finite sample performance of the proposed method is also investigated by simulation study and is shown to be much better than the kernel density estimate but close to the parametric ones.
\end{abstract}


%\titlerunning{Multivariate Bernstein Polynomial Model}        % if too long for running head

%%%%%%%%%%%%%%%%%%%%
\section{Introduction}
In nonparametric statistics, density estimation is a difficult job.  Multivariate density estimation is even more difficult.  A complete account of the multivariate density estimation can be found in the book by Scott \cite{scott2015multivariate}. The most commonly used method of multivariate density estimation is  kernel estimation. Some modifications on kernel density estimation can be found in \cite{Romano-1988, Lanh-1991, Vieu-1996}. However, the kernel method is not a maximum likelihood (ML) method. It is just a technique to smooth the discrete density corresponding to the empirical distribution by choosing appropriate bandwidth. The kernel density is actually an unbiased estimate of the convolution of the target density and the scaled kernel. Moreover, the  boundary effect of kernel estimation and the difficulty in selecting the bandwidth  still prevent the improvement upon  the accuracy of estimation.

All nonparametric estimates of  infinite-dimensional parameters such as continuous distribution  and density functions are based on models with finite-
dimensional parameter. For example, the commonly used empirical distribution is based on multinomial distribution model. Empirical likelihood method uses the same model. This model can also be viewed as a step-function approximation of the unknown population distribution function or histogram approximation of the underlying density function. However it is not a smooth approximation.

If a nonparametric model for density means any nonnegative function $f$ such that $\int f(x)dx=1$, then for any $x$ with $f(x)>0$, the information for $f(x)$ is {\em zero} \citep[see][]{Bickel-et-al-book-1998}. It was
 also showed by \cite{Ibragimov-Hasminskii-1982} that no such nonparametric model even with some smoothness assumptions
for which this information is positive. Therefore such {`nonparametric model'} is not {useful}. Box \cite{Box-1976-JASA-Sci-and-Stat} noted `{all models  are wrong, but some are useful}'. If this is agreed then such infinite dimensional {`nonparametric models'} are not even  {models} because they are {not wrong} and specify almost nothing. Therefore properly {reducing} the infinite dimensional parameter to a {finite dimensional} one is {necessary}. It is well known that in most cases the ML method gives the most efficient estimate. A working finite dimensional nonparametric density model is also necessary to apply the ML method. \added{In most cases the method of sieves \cite{Shen-and-Wong-1994}  applies.} Just like we clearly know when an estimator is parametric, with an approximate working finite dimensional nonparametric density model, one can easily  answer the question: `when is an estimator nonparametric?' \citep[see \S\S 2.4.2, 6.1.3, and 6.4 of][] {scott2015multivariate}
 \deleted{Inspired by the Bernstein \citep{Bernstein, Bernstein-1932} polynomial approximation  \citep[see][also.]{Lorentz-1963-Math-Annalen,Lorentz-1986-book-bernstein-poly}, Guan \cite{Guan-jns-2015} proposed the approximate Bernstein polynomial model for nonparametric density estimation.} \added
 {Since Vitale \citep{Vitale1975} proposed using Bernstein polynomial approximation \citep{Bernstein, Bernstein-1932,Lorentz-1963-Math-Annalen,Lorentz-1986-book-bernstein-poly} as a smoothing technique in univariate density estimation, many researches have been done to generalize it to bi- and multivariate cases including density copula \citep[see][among many others]{Tenbusch1994, Sancetta-Satchell-2004-Econ-theo}.
Inspired by these works, Guan \cite{Guan-jns-2015} proposed the approximate Bernstein polynomial model for nonparametric density estimation using ML method.} The unknown parameters contained in this model are the   coefficients and the degree of the polynomial which is also the number of unknown coefficients. The number of parameters could increase as sample size increases.
This is one of important features that characterize a nonparametric estimator \citep[see \S 2.4.2 of][]{scott2015multivariate}. The readers are refereed to \cite{Guan-jns-2015,Guan-2017-jns} for more references therein on applications of the Bernstein polynomial in estimations of density and other smooth infinite dimensional parameters. \added{Recently, \cite{Belomestny-etal-2017} proposed projection type estimation using Hermite polynomials.}  Unlike other nonparametric density estimation such as the kernel density and other applications of the Bernstein polynomial in density estimations as in \cite{Vitale1975} and \cite{Tenbusch1994}, for instance, Guan \cite{Guan-jns-2015}'s method is \replaced{an ML}{a maximum  likelihood} method using Bernstein polynomial as an approximate model \added{just like the empirical likelihood and other nonparametric ML methods} with
the degree of the polynomial together with the coefficients as a finite dimensional parameter. \added{To the authors' knowledge, all the applications of Bernstein polynomial in statistics that predate Guan \cite{Guan-jns-2015} are limited to empirically estimating the coefficients of the classical Bernstein polynomial  which are determined explicitly by the unknown density rather than the improved version of Lorentz \citep{Lorentz-1963-Math-Annalen}.  Consequently, those methods cannot take the advantage of much better degree of approximation that the improved version can achieve (see the Appendix A for details). While the ML method targets the coefficients of the improved version.} It has been shown that the Bernstein
density estimation can achieve an almost parametric optimal rate of convergence.   Simulation study showed that the small sample performance of  the Bernstein polynomial density estimate is close to that of parametric one but much better than the kernel density estimate.

Multivariate density estimation is crucial in many applications of statistics. For example the Nadaraya-Watson estimator of a nonparametric regression function
requires the nonparametric multivariate density estimate. \added{The classical Bernstein polynomial rather than the improved version has also been used in estimating multivariate
distributions including copulas \cite[see][for examples]{Sancetta-Satchell-2004-Econ-theo}. } The commonly used kernel density estimator cannot take the advantage of the boundedness of the support of a density to be estimated and therefore its rate of convergence is bounded by the optimal minimax rate \citep{stone1980}. On the other hand, however, for a density on an infinite support it is not possible to obtain
reasonable estimates of the density values outside the data range without specification of the tail behaviors.  The multivariate generalization of \cite{Guan-jns-2015}'s method is desired and is anticipated
 to provide better nonparametric multivariate density estimate than the existing methods such as the kernel density estimation.

 The paper is organized as follows.
We shall give the maximum approximate Bernstein likelihood method in Section \ref{sect: methodology} and some asymptotic results in Section \ref{asymptotics}.
The proposed methods are compared with some
existing competitors through Monte Carlo experiments in Section \ref{simulation} and are illustrated by a real dataset in Section \ref{example}. The performance of the change-point method for choosing optimal degrees is also studied in Section \ref{simulation}. Further remarks and comments are given in Section\ref{concluding remark}. The proofs of the theoretical results are relegated  to the Appendix.

\section{Methodology}\label{sect: methodology}
\subsection{Notations}
In this section
we first give some notations and definitions that will be used in the following sections.
Throughout the paper, we use bold face letters to denote vectors. For example, $\bm x = (x_1, \ldots, x_d)^\tr$ is a $d$-dimensional vector. Inequality $\bm x \le \bm y$ is understood componentwise, i.e., $x_j \le y_j$ for all $j = 1, \ldots , d$. The strict inequality
$\bm x < \bm y$ means $\bm x\le \bm y$ but $\bm x\ne \bm y$. We denote the taxicab norm by $|\bm x|=\sum_{i=1}^d |x_i|$.
 Let  $C^{(k)}[0,1]^d$ denote the class of functions $f$ on $[0,1]^d$ that have continuous partial derivatives
$f^{(\bm l)}(\bm t)\equiv\partial^{\langle\bm l\rangle} f(\bm t)/\partial t_1^{l_1}\cdots\partial t_d^{l_d}$, where $0\le \langle\bm l\rangle \equiv l_1+\cdots+l_d \le k$.

  The density of beta distribution with shape parameters $(i+1, m-i+1)$ is
$$\beta_{mi}(t)=(m+1) {m\choose i}t^{i}(1-t)^{m-i},\quad i=0,\ldots,m;\; 0\le t\le 1.$$
Then the generalized multivariate Bernstein polynomial, the multivariate polynomial with positive coefficients, can be  defined as
\begin{equation}\label{eq: d-dim ppc}
  P_{\bm m}(\bm t)=  \sum_{\bm i=0}^{\bm m}
a(\bm i) \cdot {\beta}_{\bm m \bm i}(\bm t),\quad a(\bm i)\ge 0,
\end{equation}
where $\bm t=(t_1,\ldots,t_d)$, $\bm m=(m_1,\ldots,m_d)$, $\bm i=(i_1,\ldots,i_d)$, $\sum_{\bm i=0}^{\bm m}=\sum_{i_1=0}^{m_1}\cdots \sum_{i_d=0}^{m_d}$ and ${\beta}_{\bm m \bm i}(\bm t)= \prod_{j=1}^d\beta_{m_ji_j}(t_{j}).$  The maximum number of nonzero coefficients is $K=\prod_{j=1}^d(m_j+1)$.
\subsection{Maximum Approximate Bernstein Likelihood Estimation}\label{sect: mable}
Let $\bm X=(X_1,\ldots,X_{d})^\tr$  be a $d$-dimensional random vector having a continuous joint distribution $F$ and a density $f$  with support inside the hypercube $[0, 1]^d$.
The part (iii) of Lemma \ref{thm: generalization of Thm 1 of Lorentz 1963} in the Appendix implies that we can
model the multivariate  density $f$ %with support $[0,1]^d$
approximately by
\begin{equation}\label{eq: approx bernstein poly model}
    f_{\bm m}(\bm t; \bm p)=\sum_{\bm i=0}^{\bm m}
p(\bm i) \cdot {\beta}_{\bm m \bm i}(\bm t),
\end{equation}
where $\bm p=\bm p_{\bm m} =\{p(\bm i):\; \bm 0\le \bm i\le \bm m\} \in \mathbb{S}_{\bm m}\equiv \{p(\bm i)\,:\; \bm 0\le \bm i\le \bm m,\,  p(\bm i)\ge 0,\;\;  \sum_{\bm i=0}^{\bm m} p(\bm i)=1\}$,
the $(K-1)$-simplex.
Then
  $f_{\bm m}$ is a mixture density of $K$ multiple beta distributions. Moreover, the marginal densities are also mixtures  of (multiple) beta densities. The joint cdf $F$ can be approximated by
\begin{equation}\label{eq: bernstein poly cdf}
    F_{\bm m}(\bm t; \bm p)=\sum_{\bm i=0}^{\bm m}
p(i_1,\ldots,i_d) \cdot {B}_{\bm m \bm i}(\bm t),
\end{equation}
where ${B}_{\bm m \bm i}(\bm t)= \prod_{j=1}^d B_{m_ji_j}(t_j)$ and $B_{mi}(t)$ is the cumulative distribution function of beta$(i+1, m-i+1)$, $i=0,\ldots,m$.

  Let $\bm x_i=(x_{1i},\ldots,x_{di})^\tr$, $i=1,\ldots,n$, be a sample of size $n$ from $F$.
  We assume that  $\bm p_{\bm m}$ is arranged in the  lexicographical order of $\bm i=(i_1,\ldots,i_d)$ so that $\bm p_{\bm m}$ can be treated as
   a $K$-dimensional vector.  We can define the approximate Bernstein log-likelihood
\begin{equation}\label{eq: approx loglikelihood}
    \ell(\bm p_{\bm m})=\sum_{k=1}^n \log f_{\bm m}(\bm x_k; \bm p)=\sum_{k=1}^n \log\Big\{\mathop{\sum}_{\bm i=0}^{\bm m}p(\bm i)  \,
    {\beta}_{\bm m \bm i}(\bm x_k)\Big\}.
\end{equation}
It is easy to see that if $n\ge K-1$ then $\ell(\bm p_{\bm m})$, as a function of $\bm p_{\bm m}$, is strictly concave with probability one.
The maximizer $\hat{\bm p}_{\bm m}$ of $\ell(\bm p_{\bm m})$ subject to constraint $\bm p_{\bm m}\in\mathbb{S}_{\bm m}$ is called the maximum approximate Bernstein likelihood estimate (MABLE) of $\bm p_{\bm m}$. We can estimate the underlying density and distribution functions, respectively, by the maximum approximate Bernstein likelihood estimators (MABLEs)
$\hat f_\B(\bm t)=f_{\bm m}(\bm t; \hat{\bm p})=\sum_{\bm i= \bm 0}^{\bm m}\hat p(\bm i) \,
    {\beta}_{\bm m \bm i}(\bm t)$ and
    $\hat F_\B(\bm t)
=F_{\bm m}(\bm t; \hat{\bm p})=\sum_{\bm i= \bm 0}^{\bm m}\hat p(\bm i)\,
    {B}_{\bm m \bm i}(\bm t)$.
\subsection{Optimal Degrees}\label{optimal degrees}
Starting with an initial value $\bm p^{(0)}_{\bm m}$, %(i_1,\ldots,i_d)=1/\{\prod_{k=1}^d(m_k+1)\}$,
one can use the following iteration to find the maximum likelihood estimate of $\bm p_{\bm m}$  for any given $\bm m$:

\begin{eqnarray}\label{eq: EM iteration for p}
p^{(s+1)}(l_1,\ldots,l_d)&=&
\frac{1}{n}\sum_{j=1}^{n}\frac{p^{(s)}(l_1,\ldots,l_d)\prod_{v=1}^d\beta_{m_vl_v}(x_{vj})}{\sum_{\bm i=0}^{\bm m} p^{(s)}(i_1,\ldots,i_d)\prod_{v=1}^d\beta_{m_vi_v}(x_{vj})},\\\nonumber\\\nonumber
&&
\quad {0\le l_v\le m_v}; \; \; {1\le v\le d};\;\; s=0,1,\ldots.
\end{eqnarray}
It follows from Theorem 4.2 of \cite{Redner-Walker-1984-siam} that for each $\bm m$, as $s\to\infty$, $\bm p^{(s)}_{\bm m}$ converges to
$\hat{\bm p}_{\bm m}$.

Because the marginal density of $X_j$ can be approximated by a mixture of the $(m_j+1)$ beta densities, beta$(i+1,m_j-i+1)$, $i=0,\ldots,m_j$, \cite{Guan-jns-2015} gives a lower bound for $m_j$ which is
$m_{bj}=\max\{1, \lceil\mu_j(1-\mu_j)/\sigma^2_j-3\rceil \},$
where $\mu_j=\E(X_j)$ and $\sigma_j^2=\Var(X_j)$. One can estimate $\mu_j$ and $\sigma_j^2$, respectively, by
 $\hat\mu_j=\bar x_{j\cdot}=\frac{1}{n}\sum_{i=1}^n x_{ji}$, $\hat\sigma_j^2=s_j^2=\frac{1}{n-1}\sum_{i=1}^n (x_{ji}-\bar x_{j\cdot})^2.$
We can select the optimal $m_j$ using the change-point method of \cite{Guan-jns-2015} for each $j$.
Let $\mathfrak{M}_j=\{m_{ji}=m_{j0}+i, i=0,1,\ldots,k_j\}$.
We fit the marginal data $x_{j1},\ldots,x_{jn}$, with the Bernstein model of degree $m_{ji}\in \mathfrak{M}_j$ to obtain the profile log-likelihood $\ell_{ji}=\ell_j(m_{ji})$. Let
$y_{ji}=\ell_{ji}-\ell_{j,i-1}$, $i=1,\ldots,k_j$.
We heuristically  assume that  $y_{j1},\ldots,y_{jn}$ are exponentials with  a change point  $\tau_j$  and that $m_{j\tau_j}$ is the optimal degree. We use the change-point detection method \citep[see Section 1.4 of][]{Csorgo1997a} for exponential
model to find a change-point estimate $\hat\tau_j=\arg\max_{1\le \tau\le k_j}\{R_j(\tau)\}$, where  the
likelihood ratio of $\tau$ is
$$R_j(\tau)=-\tau\log\left(\frac{\ell_{j\tau}-\ell_{j0}}{\tau}\right)-(k_j-\tau)\log\left(\frac{\ell_{jk}-\ell_{j\tau}}{k_j-\tau}\right)
+k_j\log\left(\frac{\ell_{jk_j}-\ell_{j0}}{k_j}\right),$$
for $\tau=1,\ldots,k_j$.Then we estimate the optimal $m_j$ by $\hat m_j=m_{j\hat \tau_j}$. In case $R_j(\tau)$ has multiple maximizers, we choose the smallest one as $\hat\tau_j$.

\added{For Bernstein copula the optimal $m$ as a smoothing factor was chosen to minimize the mean square error (MSE) of the density by \cite{Sancetta-Satchell-2004-Econ-theo}.
\cite{Burda-and-Prokhorov-2014-jma} proposed an alternative method to choose degree $m$ when Bernstein polynomial is used to construct prior for Bayesia multivariate infinite Gaussian mixture model.
 Other methods for model selection have been developed and extensively studied.
It seems possible to find an appropriate penalty like  AIC \cite{Akaike-AIC-1973} and BIC \cite{Schwarz-1978-aos} for choosing $m$. Unless we can find an asymptotic relationship between $m$ and some optimality criterion in terms of sample size, calculations of $\hat{\bm p}_{\bm m}$ for candidate $\bm m$'s are inevitable. }

 \subsection{Multivariate  Distribution and Density Functions Estimation}\label{support other than unit cube}
Let $\bm Y=(Y_1,\ldots,Y_{d})^\tr$  be a $d$-dimensional random vector having a continuous joint distribution $G$ and a density $g$  with support $[\bm a,\bm b]=[a_1,b_1]\times \cdots \times [a_d, b_d]$.  We transform  $\bm Y$ to %$\bm X$ as
$\bm X=(X_1,\ldots,X_d)^\tr$, where $X_i=(Y_i-a_i)/(b_i-a_i)$, $i=1,\ldots,d$. Then the distribution and density functions of  $\bm X$ are, respectively, $F(\bm x)=F(x_1,\ldots,x_d)=G\{(\bm b-\bm a)\bm x+\bm a\}$ and
$f(\bm x)= \prod_{i=1}^d(b_i-a_i) g\{(\bm b-\bm a)\bm x+\bm a\}$, where
$(\bm b-\bm a)\bm x+\bm a=\{(b_1-a_1)x_1+a_1,\ldots, (b_d-a_d)x_d+a_d\}$.  Let $\bm y_j=(y_{1j},\ldots,y_{dj})^\tr$, $j=1,\ldots,n$, be a sample from $G$. We transform the data
 to $\bm x_j=(x_{1j},\ldots, x_{dj})^\tr$, with $x_{ij}=(y_{ij}-a_i)/(b_i-a_i)$, $j=1,\ldots,n$, $i=1,\ldots,d$.  Since $f$ is a continuous $d$-variate density on the hypercube $[0,1]^d$, we can fit the transformed data $\bm x_j$, $j=1,\ldots,n$, by the Bernstein polynomial model to get the maximum likelihood estimate $\hat{\bm p}$. Then we can estimate
$g$ and $G$ respectively by
\begin{eqnarray}\label{eq: bernstein poly estimate of g}
    \hat g_\B(\bm y)&=&\frac{1}{\prod_{i=1}^d(b_i-a_i)}\sum_{\bm i=0}^{\bm m}
\hat p(i_1,\ldots,i_d) \cdot\prod_{j=1}^d\beta_{m_ji_j}\left(\frac{y_j-a_j}{b_j-a_j}\right),
\\
\label{eq: bernstein poly estimate of G}
    \hat G_\B(\bm y)&=&\sum_{\bm i=0}^{\bm m}
\hat p(i_1,\ldots,i_d) \cdot\prod_{j=1}^dB_{m_ji_j}\left(\frac{y_j-a_j}{b_j-a_j}\right).
\end{eqnarray}
\section{Asymptotic Results}\label{asymptotics}
In order to prove our asymptotic results we need the following assumption:
\begin{assump}\label{assump A1}
For each $\bm m$ large enough, there exist  a  $\bm p_0\in \mathbb{S}_{\bm m}$ and $k>0$ such that,
uniformly in $\bm t\in(0,1)^d$,
$$\frac{|f_{\bm m}(\bm t;\bm p_0)-f(\bm t)|}{f(\bm t)}\le C(d,f) m_0^{-k/2},$$
where  $m_0=\min_{1\le j\le d}m_j$, and  $C(d,f)$  depends  on $d$ and $f$ but independent of $\bm m$.
\end{assump}
 A function $f$ is said to be {\em $\gamma$--H\"{o}lder
continuous} with $\gamma\in(0,1]$ if $|f(x)-f(y)|\le C|x-y|^\gamma$ for some constant $C>0$.
By  Lemma \ref{thm: generalization of Thm 1 of Lorentz 1963} and Remark \ref{remark: allowing f to vanish along edges of the hypercube} we have
the following sufficient condition for assumption \ref{assump A1} to hold.
\begin{lemma} \label{lemma: A suuficient condition of Assump (A.1)}
Suppose that $f_0\in C^{(r)}[0,1]^d$, $r\ge0 $, $f_0(\bm t)\ge\delta_0 > 0$, and when $\langle\bm l\rangle=r$ all $f_0^{(\bm l)}$ are $\alpha$-H\"{o}lder continuous, $\alpha\in(0,1]$. If $f(\bm t)=f_0(\bm t)\prod_{i=1}^dt_i^{a_i}(1-t_i)^{b_i}$, where    $a_i$'s and $b_i$'s are nonnegative integers,
  then
assumption \ref{assump A1} is true with $k=r+\alpha$.
\end{lemma}
\begin{rem}
We have to note that  the above sufficient condition just like the smoothness conditions as people usually used are difficult to check in practise. Besides the smoothness the above sufficient condition basically allows $f$ to vanish only along the boarder of $[0,1]^d$ with zeros of integer multiplicities. In many applications we have $f=f_0$ especially when we truncate an $f$ on infinite support. In such cases $f$ does not vanish on $[0,1]^d$.
\end{rem}
Intuitively,  assumption \ref{assump A1} suggests that most of sample data can be viewed as if they were from $f_{\bm m}(\bm t;\bm p_0)$ \citep{Guan-2017-jns}. So $\ell(\bm p)$ is the likelihood of $\bm x_1,\ldots,\bm x_n$ which can be viewed as a slightly  contaminated sample from $f_{\bm m}(\bm t;\bm p_0)$.   Hence $f_{\bm m}(\bm t;\hat{\bm p})$ approximately targets at $f_{\bm m}(\bm t;\bm p_0)$ which is an approximation of $f$ satisfying assumption \ref{assump A1}.
For a given $\bm p\in \mathbb{S}_{\bm m}$, we define the  $\chi^2$-divergence ($\chi^2$-distance)
$$D^2(\bm p) =\int_{[0,1]^d}\frac{\{f_m(\bm t; \bm p)-f(\bm t)\}^2}{f(\bm t)}d\bm t
\equiv \int_{[0,1]^d}\left[\frac{f_m(\bm t; \bm p)}{f(\bm t)}-1\right]^2f(\bm t)d\bm t.$$
For density estimation we have the following result.
\begin{theorem}\label{thm: convergence rate for raw data}
Under assumption \ref{assump A1} for some positive integer $k$,  and %$\bm m={\cal O}(n^{1/k})$.
$m_0={\cal O}(n^{1/k})$.
   As $n \to\infty $, with probability one  the  maximum
value of $\ell(\bm p)$ is attained by some $\hat{\bm p}$ in the interior of
$\mathbb{B}_{\bm m}(r_n)=\{\bm p\in \mathbb{S}_{\bm m}\,:\,D^2(\bm p) \le r_n^2\}$, %n^{-2r}\}$,
where $r_n^2=\log n/n$.
Consequently we have % there is a positive constant  $C$   such that
\begin{eqnarray}\label{eq: weighted mise for raw data}
\E\int_{[0,1]^d}\frac{\{f_{\bm m}(\bm t; \hat{\bm p})-f(\bm t)\}^2}{f(\bm t)}d\bm t
&\le& % C_1 r_n^2+C_2 m^{-k-\alpha/2}\le
  \frac{\log n}{n}.
\end{eqnarray}
Because $f$ is bounded  there is a positive constant $C$ such that
\begin{eqnarray}\label{eq: mise for raw data}
\mathrm{MISE}(\hat f_\B)&=&\E \int_{[0,1]^d}\{f_{\bm m}(\bm t; \hat{\bm p})-f(\bm t)\}^2d\bm t \le %C r_n^2=
C\frac{\log n}{n}.
\end{eqnarray}
\end{theorem}
\begin{rem} The result (\ref{eq: weighted mise for raw data}) is a stronger result than (\ref{eq: mise for raw data}) because $f$ can be arbitrarily small.  The rate (\ref{eq: mise for raw data}) is an almost parametric rate of convergence for MISE. This rate can be attained by kernel type estimators for analytic densities \citep{Stepanova-2013-math-methods-stats}. It is interesting to investigate the properties of the proposed method for analytic density
functions.
Guan \cite{Guan-jns-2015} showed a similar result when $d=1$ under another set of conditions. The best  rate is ${\cal O}(n^{-1})$ that can be attained by the parametric density estimate under some regularity conditions.
\end{rem}
\begin{rem}
The Remark 1 of
\cite{juditsky2004} mentioned a  minimax rate of $\mathcal{O}(n^{-2k/(2k+1)})$ for a larger H\"{o}lder class  of order $k$ of univariate density functions even with restriction to $[0,1]$ \citep[see][also]{Ibragimov-Hasminskii-1981-book}. This  does not contradict our result because we consider a smaller class of density functions which satisfy assumption \ref{assump A1}, while a H\"{o}lder class density does not necessrily fulfill this assumption. For example, $f(x)=[x(1-x)]^{r+\alpha}/B(a,b)$, the density of beta distribution with shapes $a=b=r+1+\alpha$, where $r$ is nonnegative integer and $0<\alpha<1$. This is a member of H\"{o}lder class  of order $k=r+\alpha$ but does not satisfy assumption \ref{assump A1}.

\end{rem}
\begin{rem}\label{remark on convergence rate and choice of m}
\added{
    The Chung--Smirnov consistency rates which is a little better than (\ref{eq: mise for raw data}) are given in \cite{Leblanc-2009-JNS} for Bernstein estimators of
              distribution functions and in \cite{Janssen-etal-2012-jspi, Janssen-etal-2014-jma} for the empirical Bernstein copula  \citep{Sancetta-Satchell-2004-Econ-theo}  using some optimal choice of the degree $m$ as smoothing factor. Again these results are based on the classical Bernstein polynomial rather than the improved version of \cite{Lorentz-1963-Math-Annalen}. Therefore the degree $m$ is required to approach infinity as the sample size $n$ at a speed independent of the smoothness of the underlying density.  The proposed method of this paper presents a data-based choice
               of $m$ which can prevent overfitting problem cause by a too large $m$. It is an interesting project to improve the logarithmic factor $\log n$ to $\log\log n$ as those in the Chung--Smirnov consistency rates.
}
\end{rem}
\begin{rem}
For a density $g$ on $[\bm a,\bm b]$ with volume $V_d=\prod_{i=1}^d(b_i-a_i)$, if $g\le C_0$, then the transformed density $f$ on $[0,1]^d$ has upper bound $C=V_d C_0$. By (\ref{eq: mise for raw data}) we have
$\mathrm{MISE}(\hat g_\B)\le C_0\log n/n$. Thus the bound (\ref{eq: mise for raw data}) is not affected by $V_d$. For the supremum norm we have  $\E \max_{\bm t\in[0,1]^d}|f_{\bm m}(\bm t; \hat{\bm p})-f(\bm t)|  \le
\sqrt{C {\log n}/{n}}$.
\end{rem}

\section{Simulation Study}\label{simulation}
Consider the two-dimensional random vector $(X_1,X_2)^\tr$ with pdf $f(x_1,\,x_2)$ and support $[0,\,1]^2$.
We generate random samples of size $n$ from some distributions.
Simulation results on the estimated optimal model

degrees $\hat {\bm m}$ and mean integrated squared errors  ($\times 100$) of the density estimates based 1000 Monte Carlo runs are given in Table \ref{tbl: simulation using optimal degree m}. In this table,  $\hat f_\B$ represents the maximum approximate Bernstein likelihood density estimate; $\hat f_\K$ the kernel density computed by R package ``\textsf{ks}''. Samples of size $n$ are generated from the following distributions.
\begin{itemize}
  \item [(i)] {Beta}: joint beta distribution with pdf $f(x_1,x_2)=beta(x_1; 7,7)beta(x_2; 5,5).$
   \item [(ii)] {Normal}: bivariate normal with mean $\bm\mu=(0.5,0.5)^\tr$ and  covariance matrix $\bm\Sigma=0.125^2{1 \, 0.1 \choose 0.1 \, 1} $.

  \item [(iii)] MN: {bivariate normal mixture}  $w_1N(\bm\mu_1, \bm\Sigma_1)+w_2N(\bm\mu_2, \bm\Sigma_2)$ with parameters $w=(w_1,\, w_2)=(0.3, 0.7)$, $\bm\mu_1=(0.3, \, 0.3)^\tr$, $\bm\mu_2=(0.7, \, 0.7)^\tr, $ $$\,\bm\Sigma_1={0.065^2 \quad 0 \choose 0 \quad 0.065^2},\, \bm\Sigma_2={0.065^2 \quad 0 \choose 0 \quad 0.065^2}. $$
  \item [(iv)] {P(8,8)}: power distribution with pdf $f(x_1,x_2)=4.5(x_1^8+x_2^8)$, $0\le x_1,x_2\le 1$.
\end{itemize}

\begin{table}
  \caption{Simulation using optimal degree $m$.
  Simulation results on estimated optimal model degree $\hat {\bm m}$ and mean integrated square errors  ($\times 100$) of density estimates based 1000 Monte Carlo runs.   $\hat f_{\mathrm{B}}$: maximum Bernstein likelihood density estimate; $\hat f_{\mathrm{K}}$: kernel density. }\label{tbl: simulation using optimal degree m}
\centering
%{\footnotesize{
\begin{tabular}{*{5}{c}}%{ccccccc}
  \hline
 &	\Big. Beta	 	&	Normal	&  MN &	P(8,8)	 	 \\\cline{1-5}
		\multicolumn{5}{c}{$n=20$}	\\
$\E(\hat {\bm m})$&	(10.90,	8.72)	&	(11.50,	11.55)	 & (43.83, 44.17)&	 (~6.25,	 ~6.26)	 \\
$\Sd(\hat {\bm m})$&	(3.04,	3.08)		&	(3.10,	2.99) 	& (12.29, 12.53) &	 (2.51,	 2.54)	 \\
MISE($\hat f_{\mathrm{B}}$)&	~30.438 	 	&	~44.814 	& 151.846 &	 153.602   	\\
MISE($\hat f_{\mathrm{K}}$)&	~63.957 	 	&	~86.723 	& 217.632 &	 206.481 \\\hline
		\multicolumn{5}{c}{$n=50$}	\\
$\E(\hat {\bm m})$&	(11.42,	8.66) 		&	(12.29,	12.25) 	& (46.05,44.80) &	 (5.86,	 5.92) 	 	\\
$\Sd(\hat {\bm m})$&	(2.41,	2.26) 	&	(2.53,	2.55)	& (10.54,9.95) &	 (~1.99,	 ~2.09) \\
MISE($\hat f_{\mathrm{B}}$)&	~13.812 	&	~23.427 	& 75.429 &	 ~79.496 	\\
MISE($\hat f_{\mathrm{K}}$)&	~32.680  	&	~43.904 	& 127.248 &	 ~177.962 	 \\\hline
		\multicolumn{5}{c}{$n=100$}	\\
$\E(\hat {\bm m})$&	(11.79,	8.58) 		&	(12.87,	12.96)	& (46.68, 46.27) &	 (5.36	 5.31) 	 \\
$\Sd(\hat {\bm m})$&	(2.04,	1.79) 	&	(~2.20,	~2.23)	 & (8.13, 8.18) &	(1.24,	 1.25)	 \\
MISE($\hat f_{\mathrm{B}}$)&	~~7.403	 	&	~14.037 	& 43.340 &	 ~49.420  	\\
MISE($\hat f_{\mathrm{K}}$)&	~20.238 	 	&	~27.691 	&	82.454 & 158.145 	 \\\hline
		\multicolumn{5}{c}{$n=200$}	\\
$\E(\hat {\bm m})$&	(11.99,	8.46)	&	(13.22,	13.13)	 & (48.24, 48.08) &	 (5.95,	 5.92) 	 \\
$\Sd(\hat {\bm m})$&	(1.56,	1.48)		&	(~1.87,	~1.77) & (6.85, 6.63)	 &	 (1.33,	 1.33) 	 \\
MISE($\hat f_{\mathrm{B}}$)&	~~3.733 	 	&	~~8.884 	& 24.407 &	 ~33.672 	\\
MISE($\hat f_{\mathrm{K}}$)&	~~12.943 	 	&	~17.453 	& 53.237 &	 141.669\\\hline
%  \hline
\end{tabular}%}
\end{table}
From Table \ref{tbl: simulation using optimal degree m} we observe the following. The change-point method for choosing optimal degrees seems to give
consistent estimate of $\bm m$ for beta distribution \added{when a true $m$ exists}. For non-polynomial density distributions the optimal degrees seems increase slowly as sample size $n$ with decreasing standard deviation. The proposed density estimate $\hat f_{\mathrm{B}}$ could be 3 times and at least 1.3 times more efficient than the kernel density $\hat f_{\mathrm{K}}$. The relative efficiency of $\hat f_{\mathrm{B}}$ to $\hat f_{\mathrm{K}}$ seems increases as $n$.

\section{Real Data Application}
\label{example}
The joint density of duration $Y_1$ (in minutes) of eruptions and the waiting time $Y_2$ (in minutes) of the Old Faithful is bimodal. Based on
the data set containing $n=272$ observations which are contained in \cite{Hardle-book-1991} and also in \cite{Vena:Ripl:mode:1994}.
 Density estimation based on these data are also discussed by \cite{Silverman-1986-book} and \cite{Vena:Ripl:mode:1994}.
Petrone \cite{Petrone-1999-CJS} provides a comparison with the Baysian Bernstein density estimate.

\begin{figure}
  \includegraphics[width=5.5in]{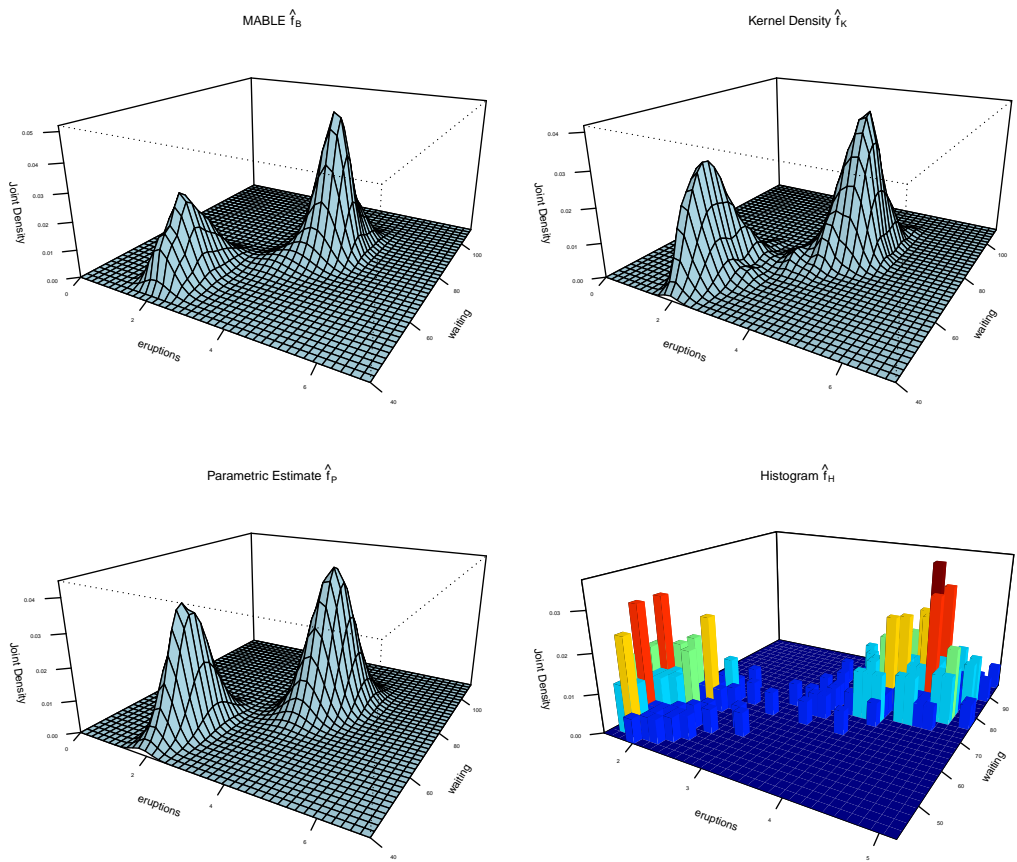}
\caption{$\hat f_\B$: the Bernstein density estimate; $\hat f_\K$: the kernel density estimate; $\hat f_{\mathrm{P}}$: the mixed normal density estimate; $\hat f_\hist$: the histogram estimate.}
\label{fig: old-faithful-density-optim-m}
\end{figure}

\begin{figure}
\centering
  \includegraphics[width=5.5in]{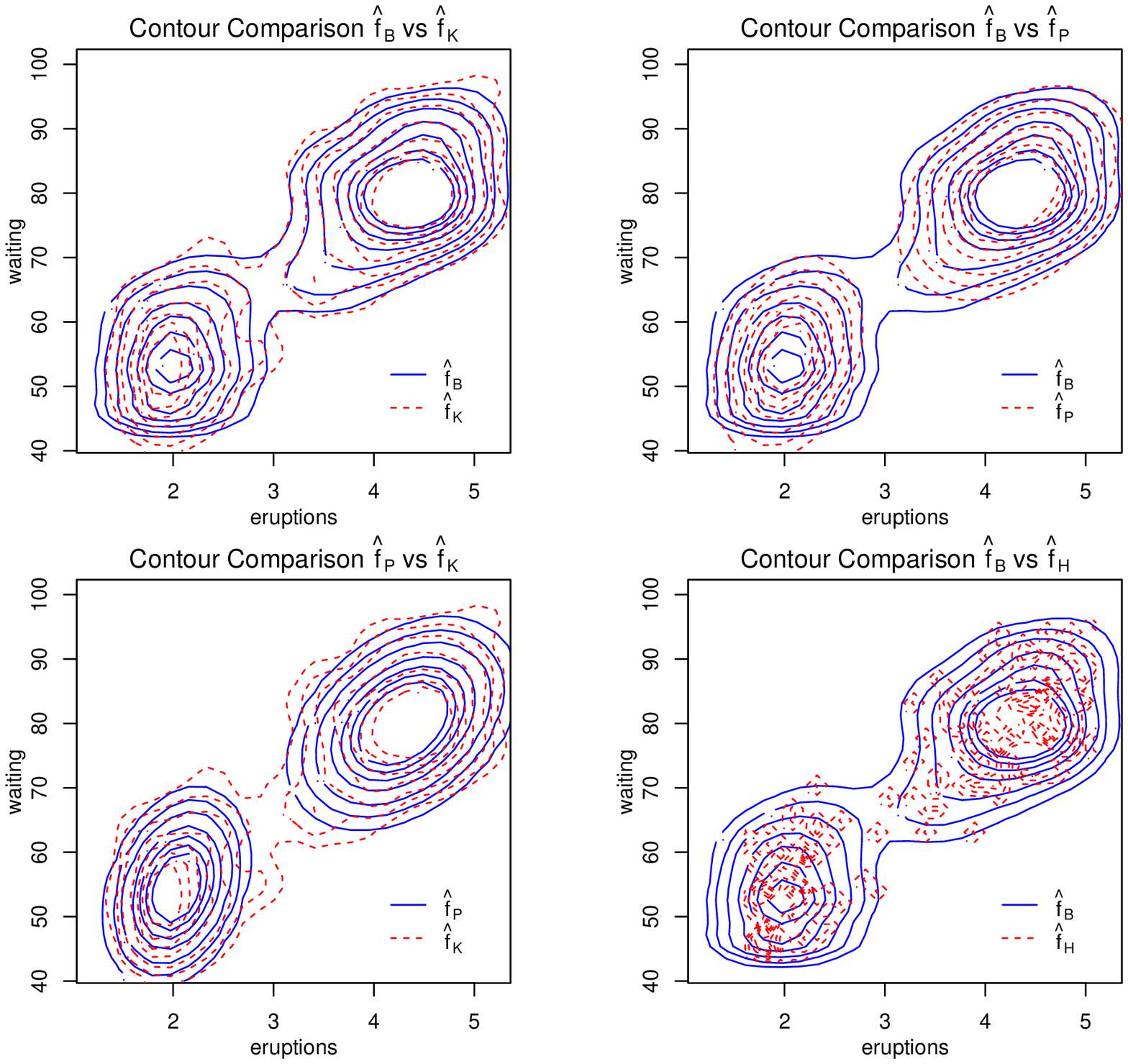}
\caption{
Pairwise contour comparison of the density estimates. $\hat f_\B$: the Bernstein density estimate; $\hat f_\K$: the kernel density estimate; $\hat f_{\mathrm{P}}$: the mixed normal density estimate; $\hat f_\hist$: the histogram estimate.}\label{fig: old-faithful-density-contour}
\end{figure}

We truncate the data by rectangle $[a_1,b_1]\times[a_2,b_2]=[0, 7]\times[0, 120]$
and transform the data to $X_i=(Y_i-a_i)/(b_i-a_i)$, $i=1,2$. Using the change-point method of \cite{Guan-jns-2015} as described in Section \ref{optimal degrees} we obtained optimal degrees $\hat m_1=95$ and $\hat m_2=88$ by fitting  the duration and waiting time data separately  with the Bernstein polynomial model.
The maximum approximate Bernstein likelihood estimate $\hat g_\B$ of the truncated density based on the data $Y_i$'s with $\hat{\bm m}=(95, 88)$ is transformed to give the
maximum approximate Bernstein likelihood estimate $\hat f_\B$ of $f$:
 $\hat f_\B(\bm x)=\hat g_\B\{(x_1-a_1)/(b_1-a_1),(x_2-a_2)/(b_2-a_2)\}/[(b_1-a_1)(b_2-a_2)].$
Figure  \ref{fig: old-faithful-density-optim-m}   shows the density estimates,  the proposed method of this paper $\hat f_{\mathrm{B}}$,  the kernel  density  $\hat f_{\mathrm{K}}$ using  R package \textsf{ks} based on mixture data, and the parametric estimate $\hat f_{\mathrm{P}}$ using R package \textsf{mixtools}, the mixture normal model, and the histogram $\hat f_{\mathrm{H}}$ of the data. Figure \ref{fig: old-faithful-density-contour}  compares the four estimates by contours. From these graphs we see that the proposed estimate is a little bit more smooth and looks more like mixture normal as many authors have assumed.
\section{Concluding Remarks}\label{concluding remark}
\added{Another point of view to look at the proposed method is that the maximum approximate Bernstein likelihood is an instance of sieve MLE \citep{Shen-and-Wong-1994} in a broad sense. \cite{Panchenko-and-Prokhorov-WP-2016} propose a sieve MLE estimator of the unknown common parameter of univariate marginal distributions in copula estimation. Here the dense subspace is indexed by $m$, an unknown parameter of the approximate model, not $n$. Because it is improper to assume $m=m(n)$ in a deterministic way it seems not easy, but still possible though, to apply the general theory like those of \citep{Shen-and-Wong-1994} to obtain or even improve the results of the present paper and even those of \cite{Guan-jns-2015}. }
Due to the many parameters to be estimated, the only drawback of the proposed method is the slow convergence of EM iteration. \added{As to computational complexity, the empirical approach and kernel type or projection estimators are clearly better than the proposed method when sample size is big. } The computation cost seems unavoidable \added{and worthy} to achieve the much better efficiency.  It is a challenge to find better algorithm to speed up the computation.  Although assumption \ref{assump A1} is not easy to check,   the sufficient condition given in Lemma \ref{lemma: A suuficient condition of Assump (A.1)} with all $a_i$'s and $b_i$'s equal to zero is fulfilled by non-vanishing densities. A nonparametric density estimator should be obtained by fitting a working nonparametric model for density. A useful and working nonparametric model should be an approximate model which contains unknown but finite number of parameters. The  method described in this is implemented in R language \citep{R-team} and will be added as a component to R package \verb"mable", maximum approximate Bernstein likelihood estimation \citep{mable}, which is available on CRAN.

\added{ Hermite polynomials and other polynomials can be used to estimate densities. However when the coefficients or other quantities that determine the polynomials are estimated using empirical distribution as in \cite{Belomestny-etal-2017} and Vitale (1975) the methods are just instance of methods of moments.   Not all polynomials can be used as a probability model. The improved version of the Bernstein polynomial as in Lorentz(1963)  with normalized coefficients happens to be a finite mixture of some specific beta densities. If an unknown underlying density has an infinite support then the tail  approaches zero at infinity at all kinds of rate. However a sample covers only a finite range. Therefore it is not possible estimate the values of the density outside that range without specifying the tail behavior. This is also the reason why we approximate a density with infinite support by a truncated one in stead of transforming it using a function, say arctan, to a density on a finite support because in  that way we still pretend to be able estimate the density outside the data range. Another advantage of the Bernstein polynomial model is that the lower degree model is nested in all higher degree ones. This makes it reasonable to use change-point method for choosing model degree. Moreover,
Hermite polynomials and many others contain infinitely many terms. For such polynomials one has to determine how many terms to used based sample. }

\added{A reviewer raised an important issue of efficiently incorporating the constraints of known marginals as in some copula situation  in the proposed estimator in finite samples. Such constraints are equivalent to the known linear combinations of $p(\bm i)$ with coefficients equal to ether beta densities or beta cdf's are known for at all $t\in[0,1]$. It will be an interesting project to choose as many as possible linearly independent constraints by selecting $t$ in [0,1] and to develop the method that can be implemented in computing algorithm.}
\section*{Acknowledgements}
The authors are grateful to the Editor and two referees for their useful comments some of which really helped to improve upon our original submission.
%Tao Wang is supported by the Natural Science Foundation of Heilongjiang Province, China(Granted No. A2017006).
%\section*{Disclosure statement}
\section*{Funding}
Tao Wang is supported by the Natural Science Foundation of Heilongjiang Province, China(Granted No. A2017006).
%
%
%% HERE WE DECLARE THE BIBLIOGRAPHYSTYLE TO USE AND THE BIBLIOGRAPHY DATABASE

%\bibliographystyle{tfnlm}
%\bibliography{bernstein,chpt,mv-ber}

\begin{thebibliography}{10}
\providecommand{\url}[1]{\normalfont{#1}}
\providecommand{\urlprefix}{Available from: }

\bibitem{scott2015multivariate}
Scott~DW. Multivariate density estimation: theory, practice, and visualization.
  John Wiley \& Sons; 2015.

\bibitem{Romano-1988}
Romano~J. On weak convergence and optimality of kernel density estimates of the
  mode. The Annals of Statistics. 1988;\hspace{0pt}16(2):629--647.

\bibitem{Lanh-1991}
Tran~LT. On multivariate variable-kernel density estimates for time series. The
  Canadian Journal of Statistics. 1988;\hspace{0pt}19(4):371--387.

\bibitem{Vieu-1996}
Vieu~P. A note on density mode estimation. Statistics \& Probability Letters.
  1996;\hspace{0pt}26(4):297--307.

\bibitem{Bickel-et-al-book-1998}
Bickel~PJ, Klaassen~CAJ, Ritov~Y, et~al. Efficient and adaptive estimation for
  semiparametric models. New York: Springer-Verlag; 1998.

\bibitem{Ibragimov-Hasminskii-1982}
Ibragimov~I, Khasminskii~R. Estimation of distribution density belonging to a
  class of entire functions. Theory of Probability \& Its Applications.
  1983;\hspace{0pt}27(3):551--562.

\bibitem{Box-1976-JASA-Sci-and-Stat}
Box~GEP. Science and statistics. Journal of the American Statistical
  Association. 1976;\hspace{0pt}71(356):791--799.

\bibitem{Shen-and-Wong-1994}
Shen~X, Wong~WH. Convergence rate of sieve estimates. Ann Statist.
  1994;\hspace{0pt}22(2):580--615.

\bibitem{Bernstein}
Bernstein~SN. D\'{e}monstration du th\'{e}or\`{e}me de {W}eierstrass fond\'{e}e
  sur le calcul des probabiliti\'{e}s. Communications of the Kharkov
  Mathematical Society. 1912;\hspace{0pt}13:1--2.

\bibitem{Bernstein-1932}
Bernstein~SN. Compl\'{e}tement \`{a} l'article de {E}. {V}oronowskaja. C R Acad
  Sci URSS. 1932;\hspace{0pt}:86--92.

\bibitem{Lorentz-1963-Math-Annalen}
Lorentz~GG. The degree of approximation by polynomials with positive
  coefficients. Mathematische Annalen. 1963;\hspace{0pt}151:239--251.

\bibitem{Lorentz-1986-book-bernstein-poly}
Lorentz~GG. Bernstein polynomials. 2nd ed. New York: Chelsea Publishing Co.;
  1986.

\bibitem{Guan-jns-2015}
Guan~Z. Efficient and robust density estimation using {B}ernstein type
  polynomials. Journal of Nonparametric Statistics.
  2016;\hspace{0pt}28(2):250--271.

\bibitem{Vitale1975}
Vitale~RA. {B}ernstein polynomial approach to density function estimation. In:
  Statistical inference and related topics (proc. summer res. inst. statist.
  inference for stochastic processes, indiana univ., bloomington, ind., 1974,
  vol. 2; dedicated to z. w. birnbaum). New York: Academic Press; 1975. p.
  87--99.

\bibitem{Tenbusch1994}
Tenbusch~A. Two-dimensional {B}ernstein polynomial density estimators. Metrika.
  1994;\hspace{0pt}41(3-4):233--253.

\bibitem{Sancetta-Satchell-2004-Econ-theo}
Sancetta~A, Satchell~S. The {B}ernstein copula and its applications to modeling
  and approximations of multivariate distributions. Econometric Theory.
  2004;\hspace{0pt}20(3):535--562.

\bibitem{Guan-2017-jns}
Guan~Z. {B}ernstein polynomial model for grouped continuous data. Journal of
  Nonparametric Statistics. 2017;\hspace{0pt}29(4):831--848.

\bibitem{Belomestny-etal-2017}
Belomestny~D, Comte~F, Genon-Catalot~V. Sobolev-hermite versus sobolev
  nonparametric density estimation on $\mathbb{R}$. Annals of the Institute of
  Statistical Mathematics. 2017;\hspace{0pt}First
  Online:https://doi.org/10.1007/s10463--017--0624--y.

\bibitem{stone1980}
Stone~CJ. Optimal rates of convergence for nonparametric estimators. The Annals
  of Statistics. 1980 11;\hspace{0pt}8(6):1348--1360.

\bibitem{Redner-Walker-1984-siam}
Redner~RA, Walker~HF. Mixture densities, maximum likelihood and the {EM}
  algorithm. SIAM Review. 1984;\hspace{0pt}26(2):195--239.

\bibitem{Csorgo1997a}
Cs\"{o}rg\H{o}~M, Horv\'{a}th~L. Limit theorems in change-point analysis. John
  Wiley \& Sons, Ltd., Chichester; 1997. Wiley Series in Probability and
  Statistics; with a foreword by David Kendall.

\bibitem{Burda-and-Prokhorov-2014-jma}
Burda~M, Prokhorov~A. Copula based factorization in {B}ayesian multivariate
  infinite mixture models. Journal of Multivariate Analysis.
  2014;\hspace{0pt}127:200--213.

\bibitem{Akaike-AIC-1973}
Akaike~H. Information theory and an extension of the maximum likelihood
  principle. In: Second {I}nternational {S}ymposium on {I}nformation {T}heory
  ({T}sahkadsor, 1971). Akad\'{e}miai Kiad\'{o}, Budapest; 1973. p. 267--281.

\bibitem{Schwarz-1978-aos}
Schwarz~G. Estimating the dimension of a model. The Annals of Statistics.
  1978;\hspace{0pt}6(2):461--464.

\bibitem{Stepanova-2013-math-methods-stats}
Stepanova~N. On estimation of analytic density functions in {$L_p$}.
  Mathematical Methods of Statistics. 2013;\hspace{0pt}22(2):114--136.

\bibitem{juditsky2004}
Juditsky~A, Lambert-Lacroix~S. On minimax density estimation on $\mathbb{R}$.
  Bernoulli. 2004 04;\hspace{0pt}10(2):187--220.

\bibitem{Ibragimov-Hasminskii-1981-book}
Ibragimov~I, Khasminskii~R. Statistical estimation. (Applications of
  Mathematics; Vol.~16). Springer-Verlag, New York-Berlin; 1981.

\bibitem{Leblanc-2009-JNS}
Leblanc~A. Chung-{S}mirnov property for {B}ernstein estimators of distribution
  functions. Journal of Nonparametric Statistics.
  2009;\hspace{0pt}21(2):133--142.

\bibitem{Janssen-etal-2012-jspi}
Janssen~P, Swanepoel~J, Veraverbeke~N. Large sample behavior of the {B}ernstein
  copula estimator. Journal of Statistical Planning and Inference.
  2012;\hspace{0pt}142(5):1189--1197.

\bibitem{Janssen-etal-2014-jma}
Janssen~P, Swanepoel~J, Veraverbeke~N. A note on the asymptotic behavior of the
  {B}ernstein estimator of the copula density. Journal of Multivariate
  Analysis. 2014;\hspace{0pt}124:480--487.

\bibitem{Hardle-book-1991}
H{\"a}rdle~W. Smoothing techniques with implementation in {S}. New York:
  Springer; 1991.

\bibitem{Vena:Ripl:mode:1994}
Venables~WN, Ripley~BD. Modern applied statistics with {S}-{P}lus. New York:
  Springer-Verlag Inc; 1994.

\bibitem{Silverman-1986-book}
Silverman~BW. Density estimation for statistics and data analysis. London:
  Chapman \& Hall; 1986. Monographs on Statistics and Applied Probability.

\bibitem{Petrone-1999-CJS}
Petrone~S. Bayesian density estimation using {B}ernstein polynomials. The
  Canadian Journal of Statistics. 1999;\hspace{0pt}27(1):105--126.

\bibitem{Panchenko-and-Prokhorov-WP-2016}
Panchenko~V, Prokhorov~A. Efficient estimation of parameters in marginals in
  semiparametric multivariate models. Concordia University, Department of
  Economics; 2011. Working Papers 11001.

\bibitem{R-team}
{R Core Team}. R: A language and environment for statistical computing. Vienna,
  Austria: R Foundation for Statistical Computing; 2018.

\bibitem{mable}
Guan~Z. mable: Maximum approximate bernstein likelihood estimation; 2018. R
  package version 1.0.

\bibitem{Hildebrandt-Schoenberg-1933}
Hildebrandt~TH, Schoenberg~IJ. On linear functional operations and the moment
  problem for a finite interval in one or several dimensions. Annals of
  Mathematics. 1933;\hspace{0pt}34(2):317--328.

\bibitem{Butzer-1953}
Butzer~PL. Linear combinations of {B}ernstein polynomials. Canadian Journal of
  Mathematics. 1953;\hspace{0pt}5:559--567.

\bibitem{Butzer-2d-Bernstein-poly-1953}
Butzer~PL. On two-dimensional {B}ernstein polynomials. Canadian Journal of
  Mathematics. 1953;\hspace{0pt}5:107--113.

\bibitem{Romanovsky-1923-bka}
Romanovsky~V. Note on the moments of a binomial $(p + q)^n$ about its mean.
  Biometrika. 1923;\hspace{0pt}15(3--4):410--412.

\bibitem{QinLawless}
Qin~J, Lawless~J. Empirical likelihood and general estimating equations. The
  Annals of Statistics. 1994;\hspace{0pt}22(1):300--325.

\end{thebibliography}
\def\cprime{$'$} \def\cprime{$'$}

 \section{Appendices}
\appendix
\section{Mathematical Preparation}\label{sect: Preliminary Results}
We denote  the modulus of continuity of function $f$ by
$\omega(f,h)=\max_{|\bm s-\bm t|<h}|f(\bm s)-f(\bm t)|$, $h>0.$
Define $\omega_{\bm r}(h)=\omega(f^{(\bm r)},h)$,   $\omega^{(r)}(h)=\max_{\langle \bm r \rangle=r}\omega_{\bm r}(h)$,
 and
 $\Delta_n=\Delta_n(t)=\max \{ {n}^{-1}, \delta_n(t) \}$,
$\delta_n=\delta_n(t)=\sqrt{{t(1-t)}/{n}}.$
If $n\le 4$ then $\Delta_n(t)=n^{-1}$ for all $t\in[0,1]$. If $n>4$ then
$$\Delta_n(t)=
\left\{
  \begin{array}{ll}
    \delta_n(t), & \hbox{$|t-0.5|\le 0.5\sqrt{1-{4}/{n}}$;} \\
    n^{-1}, & \hbox{elsewhere.}
  \end{array}
\right.$$

Let   $f$  be defined on the hypercube $[0, 1]^d$.
  The multivariate Bernstein polynomial approximation \citep{Hildebrandt-Schoenberg-1933,Butzer-1953,Butzer-2d-Bernstein-poly-1953} for $f(\bm t)$ is
\begin{eqnarray}\label{eq: d-dim Bernstein poly}
  B^f_{\bm n} (\bm t) &=& \sum_{\bm i=0}^{\bm m}
f\left({\bm i}/{\bm m}\right) \cdot b_{\bm m,\bm i}(\bm t),
\end{eqnarray}
where
 $\frac{\bm i}{\bm m} = (\frac{i_1}{m_1},\ldots, \frac{i_d}{m_d})$,
 $b_{\bm m,\bm i}(\bm t)=\prod_{j=1}^db_{m_ji_j}(t_j)$, and $b_{mi}(t)\equiv {m\choose i}t^{i}(1-t)^{m-i}$, $i=0,\ldots,m;\; 0\le t\le 1$.

The best degree of approximation by %the Bernstein polynomial
$B^f_{\bm n}(\bm t)$ is ${\cal O}(\sum_{j=1}^d m_j^{-1})$  provided that $f$ has continuous second or even higher partial derivatives.  \deleted{Most of the applications \citep{Vitale1975, Tenbusch1994} } \deleted{
of the Bernstein polynomial so far in statistics are limited to estimating   $f\left(\frac{\bm i}{\bm m}\right)$
using empirical distribution. So they are far from optimal if the underlying distribution has a smooth density function. One such example is \cite{Sancetta-Satchell-2004-Econ-theo} who used Bernstein polynomial to approximate multivariate distributions in terms of empirical Bernstein copulas.}

If $d=1$ and $f$ has a positive lower bound and higher than second order continuous derivatives \cite{Lorentz-1963-Math-Annalen} showed that there exist  better
choices of nonnegative coefficients than $f(i/m_1)$ which result in
the so called polynomial with positive coefficients and degree of approximation  better  than ${\cal O}(m_1^{-1})$.
We shall generalize the result of \cite{Lorentz-1963-Math-Annalen} for univariate polynomial with  positive coefficients to multivariate case with a little improvement.

  Let $\Lambda_{r}^{(d)}=\Lambda_{r}^{(d)}(\delta, M_0, \bm M_{r})$,
$\bm M_{r}=(M_{\bm i}=M_{i_1,\ldots,i_d},  2\le \langle\bm i\rangle \le r)$,  be the class of
 functions $f(\bm t)$ in $C^{(r)}[0,1]^d$ with the properties
$\delta\le f(\bm t)\le M_0$, $| f^{(\bm i)}(\bm t)|\le M_{\bm i}$, $\bm t\in [0,1]^d$,
for some $\delta>0$, $M_{\bm i}\ge 0$, $ 2\le \langle\bm i \rangle\le r$.
The following is an enhanced generalization of Theorem 1 of \cite{Lorentz-1963-Math-Annalen} to the multivariate positive polynomial which might be of independent interest.
\begin{lemma}\label{thm: generalization of Thm 1 of Lorentz 1963}
(i)
If   $f\in C^{(r)}[0,1]^d$, $r=0,1$, then with $C_{r,d}=d+1$
\begin{equation}\label{eq2: approx when r=0,1}
  |f(\bm x)-B^f_{\bm m}(\bm x)| \le
  C_{r,d}\omega^{(r)}(\max_{1\le j\le d}\delta_{m_j}(x_j)) \Big[\sum_{j=1}^d \delta_{m_j}(x_j)\Big]^r,\quad 0\le \bm x\le 1.
\end{equation}
(ii)
If $ r\ge 2$, $\delta>0$, $M_{\bm i}\ge 0$, be given, then there exists a constant $C_{r,d}=C_{r,d}(\delta,M_0,\bm M_{r})$ such that for each
  function $f(\bm x)\in \Lambda_{r}^{(d)}(\delta,M_0, \bm M_{r})$
one can find a sequence $P_{\bm m}(\bm x)$, $\bm m\ge 1$, of polynomials  with positive coefficients of degree $\bm m$ satisfying
\begin{equation}\label{eq2: approx of poly w pos coeff}
    |f(\bm x)-P_{\bm m}(\bm x)|\le
    C_{r,d}   \omega^{(r)}(D_{\bm m}(\bm x))D_{\bm m}^{r-2}(\bm x)  \Big[ \sum_{j=1}^d   \delta_{m_j}(x_j)\Big]^2,\quad 0\le \bm x\le 1,
\end{equation}
where $D_{\bm m}(\bm x)=\max_{1\le j\le d}\Delta_{m_j}(x_j)$.
(iii)
If $f\in \Lambda_{r}^{(d)}(\delta,M_0, \bm M_{r})$ is a  probability density function, and  $f_0^{(\bm l)}$ is $\alpha$-H\"{o}lder continuous, $\alpha\in(0,1]$,  $\langle\bm l\rangle=r$,
then normalizing the coefficients of $B^f_{\bm m}(\bm x)$ or $P_{\bm m}(\bm x)$ we obtain $f_{\bm m}(\bm x; \bm p)=\sum_{\bm i=0}^{\bm m} p(\bm i)  \cdot {\beta}_{\bm m \bm i}(\bm x)$ with coefficients  $\bm p\in \mathbb{S}_{\bm m}$ which satisfies
\begin{equation}\label{eq2: approx Bernstein model}
    |f(\bm x)-f_{\bm m}(\bm x; \bm p)|\le
    C'_{r,d}   \big(\min_{1\le i\le d}m_i\big)^{-(r+\alpha)/2},\quad 0\le \bm x\le 1,
\end{equation}
 for some constants $C'_{r,d}$.
\end{lemma}

\begin{rem}
    If $d=1$ and $r\ge 2$, then an improved version of Theorem 1 of \cite{Lorentz-1963-Math-Annalen} is
    \begin{equation}\label{eq: improved approx of poly w pos coeff}
    |f(t)-P_n(t)|\le C_r \delta_n^{2}(t)\Delta_n^{r-2}(t)\omega_r(\Delta_n(t)),\quad 0\le t\le 1,\quad n=1,\ldots.
\end{equation}
This indicates that the approximation  $P_n$ for $f$ performs especially good at the boundaries because the errors are zero at $t=0,1$. However,  results of \cite{Lorentz-1963-Math-Annalen} do not imply this when $r\ge 2$.
\end{rem}
\begin{rem}\label{remark: allowing f to vanish along edges of the hypercube}
The requirement that $f$ has a positive lower bound $\delta$ can be relaxed to allow $f$ to vanish only along the edges of $[0,1]^d$. For example,  $f(\bm x)=f_0(\bm x)\prod_{i=1}^dx_i^{a_i}(1-x_i)^{b_i}$, where $f_0\in \Lambda_{r}^{(d)}(\delta,M_0, \bm M_{r})$,  $a_i$'s and $b_i$'s are nonnegative integers. Because $P_{\bm m}(\bm x)$ is a polynomial  with positive coefficients so is $P_{\bm m}(\bm x)\prod_{i=1}^dx_i^{a_i}(1-x_i)^{b_i}$. Thus part (ii) of Lemma \ref{thm: generalization of Thm 1 of Lorentz 1963}
is still true.
\end{rem}

Using the  notations of  \cite{Lorentz-1963-Math-Annalen}, we define
  $T_{ns}(x)=\sum_{k=0}^n (k-nx)^s p_{nk}(x)$, $s=0,1,\ldots.$ It is convenient to denote $\bar T_{ns}(x)= n^{-s}T_{ns}(x)$ and $\bar T_{ns}^*(x) =n^{-s}T_{ns}^*(x):=n^{-s}\sum_{k=0}^n |k-nx|^s p_{nk}(x)$, $s=0,1,\ldots$.
In order to get a non-uniform estimate,  we need  an improved version of Lemma 1 of \cite{Lorentz-1963-Math-Annalen}:
\begin{lemma}\label{lem: estimate of Tns}
For $s\ge 0$ and some constant $A_s$
\begin{equation}\label{ineq: Tns* improved}
    \bar T_{ns}^*(x)\le A_s  \delta_n^{2\wedge s}(x) \Delta_n^{0\vee(s-2)}(x),
\end{equation}
where $a\vee b=\max(a,b)$, and $a\wedge b=\min(a,b)$. Particularly $A_0=A_1=A_2=1$, $A_3=2$ and $A_4=4$. The equality holds when $s=0,2$.
\end{lemma}
\begin{rem} Lemma 1 of \cite{Lorentz-1963-Math-Annalen} gives
%\begin{equation}\label{ineq: Tns*}
    $\bar T_{ns}^*(x)\le A_s  \Delta_n^s(x)$, $s\ge1$,
%\end{equation}
which
does not imply zero estimates  at  $x=0,1$.
\end{rem}
\begin{proof}
The special results for $s=0,1,2$ are obvious. By the formulas on P. 14 of \cite{Lorentz-1986-book-bernstein-poly}
we have
$\bar T^*_{n4}(x)=\bar T_{n4}(x)
=n^{-2}\delta_n^2(x)[3n(n-2)\delta_n^2(x)+1]\le 4 \delta_n^2(x)\Delta_n^2(x).$
By the Schwartz inequality, we have
$\bar T^*_{n3}(x)\le [\bar T^*_{n2}(x) \bar T^*_{n4}(x)]^{1/2}$ $= \delta_n(x)[\bar T^*_{n4}(x)]^{1/2} \le 2 \delta_n^2(x)\Delta_n(x).$
For $s\ge 4$,
both $T_{n,2r}(x)$ and $T_{n,2r+1}(x)$
can be expressed as
$nx(1-x)\sum_{l=0}^{r-1}[nx(1-x)]^lQ_{rl}(x),$  where $Q_{rl}(x)$ are   polynomials in $x$  with
coefficients depending on $r$ and $l$  only \citep[see Eq.5 of][]{Romanovsky-1923-bka}. Similar to \cite{Lorentz-1963-Math-Annalen}, this implies that
$\bar T_{n,2r}^*(x)=\bar T_{n,2r}(x)\le
A_{2r} \delta_n^2(x) \Delta_n^{2r-2}(x).$
By Schwartz inequality again
$\bar T^*_{n,2r+1}(x) \le [\bar T_{n2}(x)\bar T_{n,4r}(x)]^{1/2}$ $\le
A_{2r+1} \delta_n^2(x)\Delta_n^{2r-1}(x).$
The proof of the Lemma is complete.
\end{proof}
\section{Proof of Lemma \ref{thm: generalization of Thm 1 of Lorentz 1963}}

Similar to \cite{Lorentz-1963-Math-Annalen}, we want to prove that, for $r\ge 0$, there exist  polynomials of the form
\begin{equation}\label{eq2: Qnr(f)}
    Q_{\bm m  r}^f(\bm x)=\sum_{\bm k=\bm 0}^{\bm m}\biggr\{f(\sfrac{\bm k}{\bm m})\!+\!\sum_{i=2}^r \frac{1}{i!}\sum_{\langle\bm i\rangle= i}\!\!{\langle\bm i\rangle\choose \bm i}f^{(\bm i)}(\sfrac{\bm k}{\bm m})\!\!\prod_{j=1}^d \frac{1}{m_j^{i_j}}\tau_{ri_j}(x_j, m_j) \biggr\}p_{\bm m,\bm k}(\bm x),
\end{equation}
where ${\langle \bm i\rangle \choose \bm i}={\langle \bm i\rangle \choose i_1,\ldots,i_d}$ is the multinomial coefficient, and  $\tau_{ri}(x, m)$'s are polynomials, independent of $f$, in $x$ of degree $i$, in $m$ of degree $\lfloor i/2\rfloor$,
such that for each function $f\in C^{(r)}[0,1]^d$,
\begin{equation}\label{ineq2: for |f-Qnr(f)|}
    |f(\bm x)-Q_{\bm m r}^f(\bm x)|\le C'_{r,d}    \omega^{(r)}[D_{\bm m}(\bm x)] D_{\bm m}^{0\vee(r-2)}(\bm x)  \Big[ \sum_{j=1}^d   \delta_{m_j}(x_j)\Big]^{2\wedge r}
\end{equation}
with $C'_{r,d}$ depending only on $r$ and $d$.

If $f\in C^{(r)}[0,1]^d$, $r\ge1$,  by the Taylor expansion of $f(\bm k/\bm m)$ at $\bm x$, we have
$$f(\bm x)= f(\sfrac{\bm k}{\bm m})-\sum_{i=1}^r \frac{1}{i!}\sum_{\langle\bm i\rangle= i}{\langle\bm i\rangle\choose \bm i}\prod_{j=1}^d(\sfrac{k_j}{m_j}-x_j)^{i_j}
f^{(\bm i)}(\bm x)$$
$$\hspace{7em}+\frac{1}{r!}\biggr\{\sum_{\langle\bm i\rangle= r}{r\choose \bm i}\prod_{j=1}^d(\sfrac{k_j}{m_j}-x_j)^{i_j}\big[f^{(\bm i)}(\bm x) - f^{(\bm i)}(\bm \xi_{\bm k}^{(r)}) \big]\biggr\},$$
where $\bm\xi_{\bm k}^{(r)}$ is on the  line segment connecting $\bm x$ and $\bm k/\bm m$. This equation is also true when $r=0$ by defining $\bm\xi_{\bm k}^{(0)}=\bm k/\bm m$  and the empty sum to be zero. Multiplying both sides by $p_{\bm m,\bm k}(\bm x)$
and taking summation over  $\bm 0\le \bm k\le \bm m$, we obtain
\begin{equation}\label{Taylor expansion of f}
f(\bm x)=B^f(\bm x)-\sum_{i=2}^r \frac{1}{i!}\sum_{\langle\bm i\rangle= i}{\langle\bm i\rangle\choose \bm i}\prod_{j=1}^d\bar T_{m_ji_j}(x_j)f^{(\bm i)}(\bm x)+R_{\bm m}^{(r)}(\bm x),
\end{equation}
where
$r\ge0$, empty sum is zero, and
$$R_{\bm m}^{(r)}(\bm x)=\frac{1}{r!}\biggr\{\sum_{\langle\bm i\rangle= r}{r \choose \bm i}\sum_{\bm k=0}^{\bm m}\prod_{j=1}^d\frac{1}{m_j^{i_j}}\left( k_j-m_jx_j\right)^{i_j}p_{m_jk_j}(x_j)\big[f^{(\bm i)}(\bm x) - f^{(\bm i)}(\bm \xi_{\bm k}^{(r)})\big]\biggr\}.$$
For each $\delta>0$, define
$\lambda=\lambda(\bm x,\bm y;\delta)=\left\lfloor  {|\bm x-\bm y|}/{\delta}\right\rfloor$, where $\lfloor x\rfloor$ is the integer part of $x\ge 0$.
Then $\lambda \delta\le |\bm x-\bm y|<(\lambda+1)\delta$, and
 for   $g\in C[0,1]^d$,
 $|g(\bm x)-g(\bm y)|\le
(\lambda+1) \omega(g,\delta).$

If  $f\in C^{(r)}[0,1]^d$, $r=0,1$,  then similar to the proofs of Theorems 1.6.1 and 1.6.2 of \cite[][pp. 20-- 21]{Lorentz-1986-book-bernstein-poly} and by (\ref{Taylor expansion of f}) we have
$|f(\bm x)-B^f_{\bm m}(\bm x)|
=|R_m^{(r)}(\bm x)|$.
Because $\lambda(\bm x,  \bm k/\bm m;\delta)
\le \delta^{-1}\sum_{j=1}^d |k_j-m_j x_j|/m_j$, by Lemma \ref{lem: estimate of Tns} with $s=0,1,2$, we have
\begin{align}\nonumber
  |f(\bm x)-B^f_{\bm m}(\bm x)|
&\le \sum_{\langle\bm i\rangle= r}\omega_{\bm i}(\delta)\biggr[\prod_{j=1}^d \bar T^*_{m_j,i_j}(x_j)
\!\!+\!\!\frac{1}{\delta}  \sum_{l=1}^d \bar T^*_{m_l,i_l+1}(x_l)\!\!
%\\\nonumber &\hspace{8em}\cdot
\!\!\mathop{\prod_{1\le j\le d}}_{j\ne l}\!\!\bar T^*_{m_j,i_j}(x_j) \biggr]\\\label{ineq r=0,1}
&\le \sum_{\langle\bm i\rangle= r}\omega_{\bm i}(\delta)\biggr[\prod_{j=1}^d\delta_{m_j}^{i_j}(x_j)
 +\frac{1}{\delta}  \sum_{l=1}^d \delta_{m_l}^{i_l+1}(x_l)\!\!\mathop{\prod_{1\le j\le d}}_{j\ne l}\delta_{m_j}^{i_j}(x_j) \biggr].
\end{align}
The estimates in (\ref{eq2: approx when r=0,1}) follow from (\ref{ineq r=0,1})  with $\delta=\max_{1\le j\le d}\delta_{m_j}(x_j)$. This also proves  (\ref{ineq2: for |f-Qnr(f)|}) with $r=0,1$ and $Q_{\bm m r}^f=B^f$. So part (i) is proved.

If $r\ge 2$, then  we have
%\begin{small}
\begin{align*}
%  &\quad |R_{\bm n}|\\
 |R_{\bm m}^{(r)}(\bm x)|
  &\le  \frac{1}{r!}\biggr\{\sum_{\langle\bm i\rangle= r}{r \choose \bm i}\omega_{\bm i}(\delta)\biggr[\prod_{j=1}^d\bar T^*_{m_ji_j}(x_j)
 % \\ &\quad
  +\frac{1}{\delta}\sum_{l=1}^d  \bar T^*_{m_l,i_l+1}(x_l)\!\!\mathop{\prod_{1\le j\le d}}_{j\ne l}  \bar T^*_{m_ji_j}(x_j)\biggr]  \biggr\} %\\
\\
 \quad\quad &\le  \frac{1}{r!}\biggr\{\sum_{\langle\bm i\rangle= r}{r \choose \bm i}\omega_{\bm i}(\delta)\biggr[\prod_{j=1}^d A_{i_j} \delta_{m_j}^{2\wedge i_j}(x_j) \Delta_{m_j}^{0\vee (i_j-2)}(x_j)   \\
  &\quad +\frac{1}{\delta}\sum_{l=1}^d  A_{i_l+1}\delta_{m_l}^{2\wedge(i_l+1)}(x_l)\Delta_{m_l}^{0\vee(i_l-1)}(x_l)\!\!\!\mathop{\prod_{1\le j\le d}}_{j\ne l} \!\!\! A_{i_j}\delta_{m_j}^{2\wedge i_j}(x_j) \Delta_{m_j}^{0\vee (i_j-2)}(x_j)\biggr]\biggr\}.
\end{align*}
Choosing
$\delta= D_{\bm m}(\bm x)$,  we have
\begin{align*}
 |R_{\bm m}^{(r)}(\bm x)| &\le  \omega^{(r)}(\delta) \frac{1}{r!}\biggr\{\sum_{\langle\bm i\rangle= r}{r \choose \bm i} \prod_{j=1}^d A_{i_j} \delta_{m_j}^{2\wedge i_j}(x_j) \Delta_{m_j}^{0\vee (i_j-2)}(x_j)   \\
  &\quad + \sum_{l=1}^d\sum_{\langle\bm i\rangle= r}{r \choose \bm i}   A_{i_l+1}\delta_{m_l}^{2\wedge i_l}(x_l)\Delta_{m_l}^{0\vee(i_l-2)}(x_l)\!\!\!\!\mathop{\prod_{1\le j\le d}}_{j\ne l}\!\!\!\! A_{i_j}\delta_{m_j}^{2\wedge i_j}(x_j) \Delta_{m_j}^{0\vee (i_j-2)}(x_j)\biggr\}\\
  &\le   C(r,d) \omega^{(r)}(\delta)\mathop{\max}_{1\le j\le d} \Delta_{m_j}^{r-2}(x_j)  \Big[ \sum_{j=1}^d   \delta_{m_j}(x_j)\Big]^2.
\end{align*}

Similar to \cite{Lorentz-1963-Math-Annalen} we shall prove the existence of $Q_{\bm m r}^f$ by induction in $r$. Assuming that all $Q_{\bm m i}^f$ for $i<r$
are established, we iteratively define
\begin{equation}\label{eq2: iteration for Qnr(f)}
    Q_{\bm m r}^f(\bm x)= B^f(\bm x)-\sum_{i=2}^r \frac{1}{i!}\sum_{\langle\bm i\rangle= i}{\langle\bm i\rangle\choose \bm i}\prod_{j=1}^d\bar T_{m_ji_j}(x_j)  Q_{\bm m, r-i}^{f^{(\bm i)}}(\bm x).
\end{equation}
By Lemma \ref{lem: estimate of Tns} and the inductive assumption, (\ref{ineq2: for |f-Qnr(f)|}) is  satisfied by (\ref{eq2: iteration for Qnr(f)}) as following.
%\begin{small}
\begin{align*}
  |f(\bm x)-Q_{\bm m r}^f(\bm x)|
   &\le  \sum_{i=2}^r \frac{1}{i!}\sum_{\langle\bm i\rangle= i}\!\!{\langle\bm i\rangle\choose \bm i}\!\!\prod_{j=1}^d\bar T^*_{m_ji_j}(x_j) |f^{(\bm i)}(\bm x)\!-\!Q_{\bm m,r-i}^{f^{(\bm i)}}(\bm x)|+|R_{\bm m}^{(r)}(\bm x)| \\
   &\le \sum_{i=2}^r \frac{C''_{r,d}}{i!}D_{\bm m}(\bm x)^{i-2}(\bm x)\biggr[\sum_{j=1}^d\delta_{m_j}(x_j)\biggr]^{2}
\\   &\quad \cdot \omega^{(r)}[D_{\bm m}(\bm x)] D_{\bm m}^{0\vee(r-i-2)}(\bm x)\biggr[ \sum_{j=1}^d   \delta_{m_j}(x_j)\biggr]^{2\wedge(r-i)} +|R_{\bm m}^{(r)}(\bm x)|
    \\
   &\le C'''_{r,d}\omega^{(r)}[D_{\bm m}(\bm x)] D_{\bm m}^{0\vee(r-2)}(\bm x) \biggr[ \sum_{j=1}^d   \delta_{m_j}(x_j)\biggr]^{2}.
\end{align*}

Since $f(\bm x)\ge \delta>0$, by an obvious generalization of remark (a) on p. 241 of \cite{Lorentz-1963-Math-Annalen} with $\bm h=1/\bm m$ we know that $P_{\bm m+r}(\bm x)=Q_{\bm m r}^f(\bm x)$ is a $d$-variate polynomial of degree $\bm m+ r=(m_1+r,\ldots,m_d+r)$ with positive coefficients
for all $\bm m\ge \bm m_{r}(m, \mathscr M_{r})$ so that $$|f(\bm x)-P_{\bm m+r}(\bm x)|\le C_{r,d}\omega^{(r)}[D_{\bm m}(\bm x)] D_{\bm m}^{r-2}(\bm x) \biggr[ \sum_{j=1}^d   \delta_{m_j}(x_j)\biggr]^{2}. $$
Then (\ref{eq2: approx of poly w pos coeff}) follows for all $\bm m$ and a larger $C_{r,d}$ from  $\Delta_{m_j}=\mathcal{O}(\Delta_{m_j+r})$ for all $r\ge 2$. The proof of part (ii) is complete.

We omit the proof of part (iii) because it is almost the same as that of Lemma \ref{lemma: A suuficient condition of Assump (A.1)} below.

\section{Proof of Lemma \ref{lemma: A suuficient condition of Assump (A.1)}}
By parts (i) and (ii) of Lemma \ref{thm: generalization of Thm 1 of Lorentz 1963} we have
$ f_0(\bm t) -P_{\bm m}(\bm t)= R_{\bm m}(\bm t)$,
where $P_{\bm m}(\bm t)$ is a polynomial with positive coefficients and $R_{\bm m}(\bm t)$ satisfies
$|R_{\bm m}(\bm t)|\le C_0(d,f) m_0^{-(r+\alpha)/2}$, $0\le \bm t\le 1.$  So $f(\bm t)- Q_{\tilde{\bm m}}(\bm t) = R_{\tilde{\bm m}}(\bm t)$,
where   $Q_{\tilde{\bm m}}(\bm t)=P_{\bm m}(\bm t)\prod_{i=1}^dt_i^{a_i}(1-t_i)^{b_i}=\sum_{\bm i=0}^{\tilde{\bm m}}
a(\bm i) \cdot {\beta}_{\tilde{\bm m} \bm i}(\bm t)$ is  a polynomial of degree $\tilde{\bm m}=\bm m+\bm a+\bm b$ with positive coefficients,  $R_{\tilde{\bm m}}(\bm t)=R_{\bm m}(\bm t)\prod_{i=1}^dt_i^{a_i}(1-t_i)^{b_i}$, and
 $|R_{\tilde{\bm m}}(\bm t)|\le C_0(d,f) m_0^{-(r+\alpha)/2}$, $0\le \bm t\le 1.$ For large $\bm m$, $\rho_{\tilde{\bm m}}:=\int_{[0,1]^d}R_{\tilde{\bm m}}(\bm t)d\bm t
 \le C_0(d,f) m_0^{-(r+\alpha)/2}<c_0<1$.
Since $f(\bm t)$ and ${\beta}_{\tilde{\bm m} \bm i}(\bm t)$ are   densities on $[0,1]^s$, $ \sum_{\bm i=0}^{\tilde{\bm m}}
a(\bm i)=1-\rho_{\tilde{\bm m}}>0$. So normalizing $a(\bm i)$ we obtain
 $f_{\tilde{\bm m}}(\bm t;\bm p_0)=Q_{\tilde{\bm m}}(\bm t)/(1-\rho_{\tilde{\bm m}})= \sum_{\bm i=0}^{\tilde{\bm m}}
p_0(\bm i) \cdot {\beta}_{\tilde{\bm m} \bm i}(\bm t),$
where $p_0(\bm i)=a(\bm i)/(1-\rho_{\tilde{\bm m}})$. Noticing that $f_0(\bm t)\ge \delta_0>0$, we have
\begin{align*}%\label{ineq: for |f-fm|/f}
     {|f_{\tilde{\bm m}}(\bm t;\bm p_0)-f(\bm t)|}/{f(\bm t)}
 =&~
  {(1-\rho_{\tilde{\bm m}})^{-1}}\left| {R_{\tilde{\bm m}}(\bm t)}/{f(\bm t)}+\rho_{\tilde{\bm m}}\right|\\
=&~{(1-\rho_{\tilde{\bm m}})^{-1}}\left| {R_{{\bm m}}(\bm t)}/{f_0(\bm t)}+\rho_{\tilde{\bm m}}\right|\\
\le &~(1-c_0)^{-1}C_0(d,f)(1/\delta_0+1)m_0^{-(r+\alpha)/2}.
\end{align*}
The proof is complete.
\section{Proof of Theorem \ref{thm: convergence rate for raw data}}
The approximate Bernstein log likelihood is
$\ell(f_{\bm m})=\ell(\bm p)=\sum_{i=1}^n \log[f_{\bm m}(\bm x_i;\bm p)].$
Define the log-likelihood ratio
$\mathcal{R}(\bm p)=\ell(f) - \ell(\bm p),$
where  $\ell(f)=\sum_{i=1}^n\log f(\bm x_i)$.
For an $\epsilon_0\in (0,1)$, we define subset $\mathcal{A}_{\bm m}(\epsilon_0)$ of $\mathbb{S}_{\bm m}$ so that, for all $\bm x \in [0,1]^d$, $|f_{\bm m}(\bm x;\bm p)-f(\bm x)|/f(\bm x)\le \epsilon_0<1$. Clearly, such an $\mathcal{A}_{\bm m}(\epsilon_0)$ is nonempty since $\bm p_0\in \mathcal{A}_{\bm m}(\epsilon_0)$.
By Taylor expansion we have, for all $\bm p\in \mathcal{A}_{\bm m}(\epsilon_0)$,
\begin{eqnarray*}
  \mathcal{R}(\bm p)&=& -\sum_{i=1}^n \Big[Z_i(\bm p)-\frac{1}{2}Z_i^2(\bm p)\Big]+\mathcal{O}(R_{{\bm m}n}(\bm p)) \quad a.s.,
 \end{eqnarray*}
where $R_{{\bm m}n}(\bm p)=\sum_{i=1}^nZ_i^2(\bm p)$, and
$Z_i(\bm p)=[{f_{\bm m}(\bm x_i;\bm p)-f(\bm x_i)}]/{f(\bm x_i)}$, $i=1,\ldots,n.$
Since $\E[Z_i(\bm p)]=0$, $\sigma^2[Z_i(\bm p)]=\E[Z_i^2(\bm p)]=D^2(\bm p)$,
by the law of iterated logarithm  we have
$$\sum_{i=1}^n Z_i(\bm p)/\sigma[Z_i(\bm p)]=\mathcal{O}(\sqrt{n\log\log n}) \quad a.s. $$
By the strong law of large numbers we have
\begin{equation}\label{log likelihood ratio fm(p)}
\mathcal{R}(\bm p) =\frac{n}{2} D^2(\bm p) - \mathcal{O}(D(\bm p)\sqrt{n\log\log n}) +\mathcal{O}(nD^2(\bm p)) \quad a.s.
\end{equation}
If $D^2(\bm p)= r_n^2=\log n/n$, then,  by (\ref{log likelihood ratio fm(p)}),
there is an $\eta>0$ such that
 $\mathcal{R}(\bm p)\ge  %\frac{1}{2}\log n+o(\log n)\ge
\eta \log n$,  a.s..
At $\bm p=\bm p_0$, if $m_0=Cn^{1/k}$ then by assumption \ref{assump A1} we have $D^2(\bm p_0)=\mathcal{O}(m_0^{-k})=\mathcal{O}(n^{-1})$.
By (\ref{log likelihood ratio fm(p)}) again we have
 $\mathcal{R}(\bm p_0)=
\mathcal{O}(\sqrt{\log\log n})$  a.s..
Therefore, similar to the proof of Lemma 1 of \cite{QinLawless}, we have
$$D^2(\hat{\bm p})
=\int_{[0,1]^d}\frac{[f_{\bm m} (\bm x;\hat{\bm p})-f(\bm x)]^2}{f(\bm x)}d\bm x< \frac{ \log n}{n} \quad a.s. $$

The inequality
(\ref{eq: weighted mise for raw data}) follows immediately. Based on (\ref{eq: weighted mise for raw data}) and the boundedness of $f$, the inequality (\ref{eq: mise for raw data}) is obtained.

 \end{document}